\documentclass[aps,prl,twocolumn,superscriptaddress,showpacs,nobibnotes]{revtex4-1}

\usepackage{graphicx}
\usepackage{dcolumn}
\usepackage{amsmath}

\newcommand*{\etal}{\emph{et al.}}
\newcommand*{\wn}{cm$^{-1}$}
\newcommand*{\Hm}{H$_{2}$}
\newcommand*{\Dm}{D$_{2}$}
\newcommand*{\EFX}{$EF\,{}^{1}\Sigma_{g}^{+}-X\,{}^{1}\Sigma_{g}^{+}$}
\newcommand*{\XQED}{$X_\mathrm{rel+QED}(J)$}
\newcommand*{\XQEDg}{$X_\mathrm{rel+QED}(J=0)$}
\usepackage{verbatim}

\begin{document}

\title{QED effects in molecules: test on rotational quantum states of H$_2$}

\author{E. J. Salumbides}
\email[]{e.j.salumbides@vu.nl}
\affiliation{Institute for Lasers, Life and Biophotonics Amsterdam, VU University, De Boelelaan 1081, 1081 HV Amsterdam, The Netherlands}
\affiliation{Department of Physics, University of San Carlos, Cebu City 6000, Philippines}
\author{G. D. Dickenson}
\author{T. I. Ivanov}
\author{W. Ubachs}
\email[]{w.m.g.ubachs@vu.nl}
\affiliation{Institute for Lasers, Life and Biophotonics Amsterdam, VU University,
De Boelelaan 1081, 1081 HV Amsterdam, The Netherlands}

\date{\today}

\begin{abstract}
Quantum electrodynamic effects have been systematically tested in the progression of rotational quantum states in the $X\,{}^{1}\Sigma_{g}^{+}, v=0$ vibronic ground state of molecular hydrogen. High-precision Doppler-free spectroscopy of the $EF\,{}^{1}\Sigma_{g}^{+}-X\,{}^{1}\Sigma_{g}^{+}$ (0,0) band was performed with $0.005$ cm$^{-1}$ accuracy on rotationally-hot H$_2$ (with rotational quantum states $J$ up to $16$). QED and relativistic contributions to rotational level energies as high as $0.13$ cm$^{-1}$ are extracted, and are in perfect agreement with recent calculations of QED and high-order relativistic effects for the H$_2$ ground state.
\end{abstract}


\maketitle

Quantum electrodynamics (QED) has been hailed as the most successful theory in physics as its predictions are in remarkable agreement with a variety of extremely precise experiments. Precision tests of QED include free-particle systems (e.g., the anomalous magnetic moment of the electron~\cite{Hanneke2008} and of the muon~\cite{Bennett2006}) and bound atomic systems, such as the (one-electron) H atom~\cite{Biraben2009}, heavy hydrogenlike ions (e.g., U$^{91+}~$\cite{Gumberidze2005}), and the (two-electron) He atom~\cite{Kandula2010}. Recent reviews of QED theory and precision tests in simple atoms and ions can be found in Refs.~\cite{Eides2001, Karshenboim2005}.
Progress has been made in QED calculations for the smallest (one-electron) molecular ion H$_2{}^+$~\cite{Korobov2009}, and systematic high-resolution spectroscopic investigations have been proposed based on cooled ions in a trap~\cite{Koelemeij2007}. 
Recently, the problem of QED calculations (also including high-order relativistic corrections) in the smallest neutral molecule H$_2$, the benchmark system of molecular physics, has been addressed by Pachucki and co-workers.
The experimental determinations for the ionization and dissociation energies of molecular hydrogen and its isotopomers to the $\sim10^{-8}$ level of accuracy for \Hm~\cite{Liu2009}, \Dm~\cite{Liu2010} and HD~\cite{Sprecher2010,Sprecher2011} had stimulated calculations of these quantities. The \emph{ab initio} calculations of the \Hm\ and \Dm\ dissociation energies~\cite{Piszczatowski2009}, and that for HD~\cite{Pachucki2010b}, for which relativistic and QED effects for the lowest $J=0$ level were computed, are in excellent agreement with experimental results for the dissociation energies.
The QED calculations were thereupon extended using the theoretical framework in Refs.~\cite{Piszczatowski2009, Pachucki2010b}\ to the full set of rovibrational levels in the ground states of \Hm, \Dm\ and HD. When combined with the updated nonrelativistic \emph{ab initio} calculations including adiabatic and nonadiabatic effects~\cite{Pachucki2009}, a full set of QED rovibrational energies for \Hm, \Dm, and HD are now available~\cite{Komasa2011,Pachucki2010b}.

In the present investigation, we pursue a systematic study of QED effects (including higher-order relativistic and radiative effects) in a progression of 16 rotational quantum states in the $X^1\Sigma_g^+,\,v=0$ ground state of H$_2$.
For this purpose, a nonthermal \Hm\ population distribution was produced via ultraviolet(UV)-induced photodissociation of hydrogen bromide (HBr) and a subsequent chemical reaction: H + HBr $\rightarrow$ Br + H$_2$($J$) \cite{Aker1989a}, thus forming rotationally hot H$_2$($J$) with quantum states detected up to $J=16$. Here we follow similar procedures employed by Heck~\etal~\cite{Heck1995} for the production and (low-resolution) spectroscopic studies of hot \Dm.

The principle behind the derivation of the accurate rotational level energies is based on the precise laser spectroscopic measurements of $EF\,{}^{1}\Sigma_{g}^{+}-X\,{}^{1}\Sigma_{g}^{+}, Q(J)$ two-photon transitions ($Q$ lines denote $\Delta J=0$ transitions \cite{Hannemann2006}) in the range $J=6-16$, combined with the existing information on the excited states. This approach depends on the availability of accurate level energies of $EF^1\Sigma_g^+,\,(v=0,\,J=2-12)$ from studies determining $EF, J=0,1$ anchor levels~\cite{Hannemann2006,Salumbides2008} and performing Fourier-transform spectroscopic studies on the manifolds of excited states in \Hm~\cite{Bailly2010}. By measuring additional $O$-branch ($\Delta J=-2$) and $S$-branch ($\Delta J=+2$) two-photon transitions, $X\,{}^{1}\Sigma_{g}^{+}$ ground state energy splittings were derived extending the rotational sequence up to $J=16$.
Experimental relativistic and QED corrections to the ground state rotational levels are finally obtained by subtracting the calculated nonrelativistic energies, taken from the recent \emph{ab initio} study of Pachucki and Komasa~\cite{Pachucki2009}.

In the experiment, a pulsed-dye-amplifier (PDA) laser system, which is injection seeded by a continuous-wave (cw) ring dye laser~\cite{Eikema1996}, is operated at wavelengths between $610-645$ nm. Pulsed ultraviolet radiation of narrow bandwidth in the $203-215$ nm wavelength range is obtained via two-stage third-harmonic up-conversion of the PDA-output radiation in nonlinear crystals. The method of 2+1 resonance-enhanced multiphoton ionization (REMPI) is used to probe the $EF-X$ transitions~\cite{Hannemann2006}. We note that the production of hot H$_2$ and the spectroscopy is performed by the same UV laser pulse and occurs within 5 ns. Ions produced via REMPI are accelerated through a time-of-flight mass selector and detected by an imaging system composed of a multichannel plate, phosphor screen and photomultiplier, and finally digitally registered.

To obtain Doppler-free spectra, two counterpropagating UV laser beams are crossed with the molecular beam of HBr, in which rotationally hot H$_2$ is produced. The counterpropagating UV beams are aligned using the Sagnac geometry \cite{Hannemann2007}\ to avoid residual Doppler shifts, which we estimate to be less than 1 MHz for the investigated H$_2$ transitions. The application of a separate 355-nm ionization laser (pulse delayed by 30 ns with respect to the excitation laser) enables us to reduce the UV radiation intensities, in order to produce the narrowest line profiles (350 MHz) and minimize ac Stark shifts. Typically, only $50-200$~$\mu$J is used in the experiment (for hot-\Hm\ production and two-photon excitation) out of the $\sim1$~mJ UV radiation produced. For most (strong) transitions, we recorded spectra at different UV intensities for extrapolating to zero-intensity frequencies within $\sim50$~MHz. dc Stark shifts were found to be negligible.

\begin{figure}
\includegraphics[width=0.9\columnwidth]{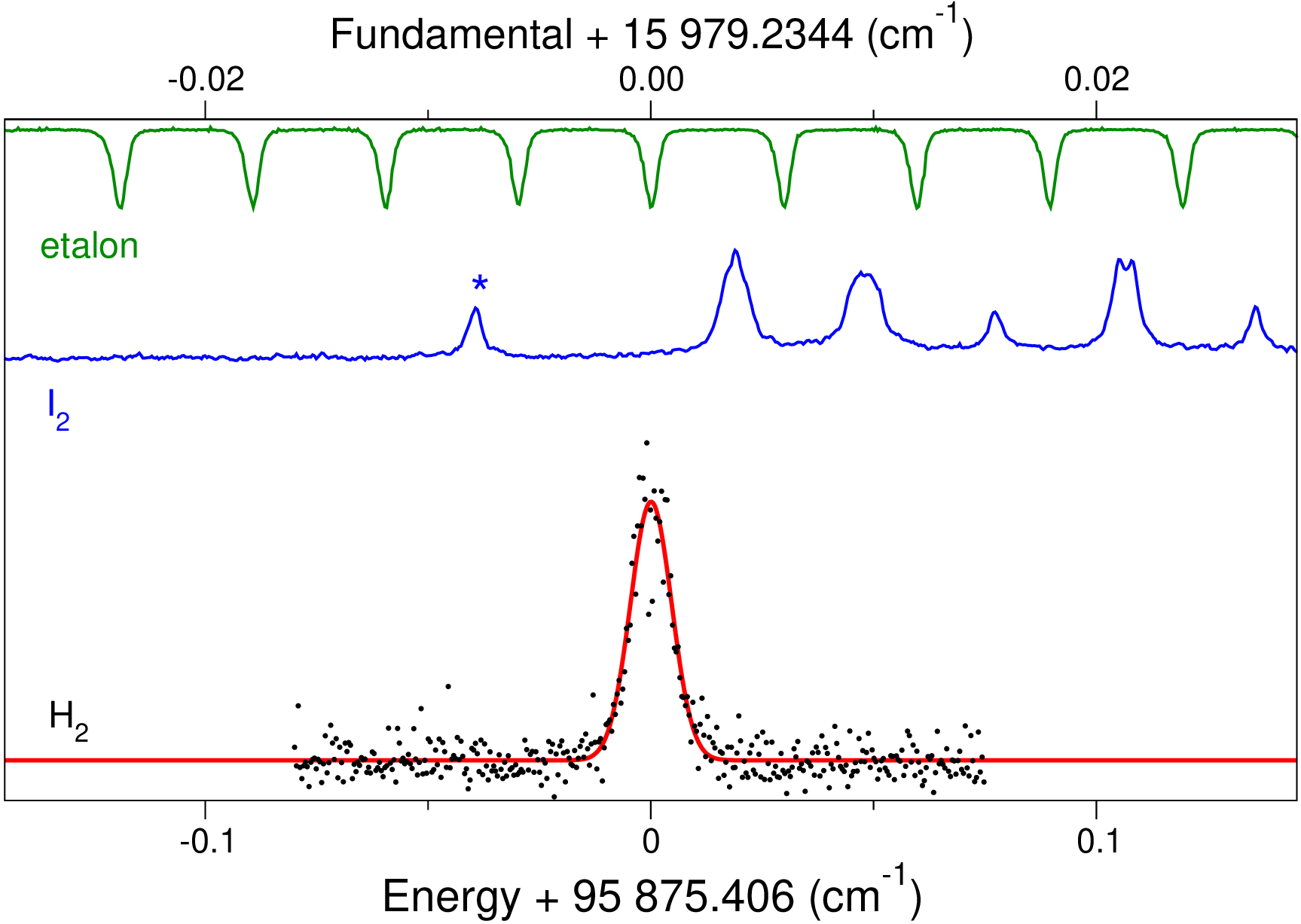}%
\caption{\label{fig:spectrum} (Color online) Recording of the \EFX\ (0,0) two-photon $O(9)$ transition. Relative frequency markers are provided by the fringes of a stabilized etalon (FSR=$148.96$ MHz), while absolute calibration is achieved using the known position~\cite{Xu2000} of an I$_2$ hyperfine component (marked by *) as the absolute frequency standard.}
\end{figure}

To provide a relative frequency scale, part of the cw-output of the ring dye laser is directed to an actively-stabilized reference etalon with free spectral range (FSR) of $148.96$~MHz. Another part of the ring laser cw-output is utilized in a Doppler-free molecular iodine (I$_2$) saturation spectroscopy setup for absolute frequency calibration~\cite{Xu2000}. A typical measurement result is depicted in Fig.~\ref{fig:spectrum}, which shows the simultaneous recording of the \Hm\ $EF-X,\,(0,0),\,O(9)$ transition, the stabilized etalon fringes, and the hyperfine-resolved I$_2$ calibration line
[$R(82)\,(10-4)$ a$_1$ component at $479\,045\,193.7(5)$ MHz].
Note that in Fig.~\ref{fig:spectrum}, the energy scale (lower axis) of the H$_2$ resonance is exactly sixfold that of the etalon markers and the I$_2$ reference spectrum (upper axis). The uncertainty contribution to the $EF-X$ transition frequencies from the scan (non)linearity and the I$_2$ calibration are estimated to be 5~MHz, respectively. For typical \Hm\ transitions recorded, the signal-to-noise (SNR) ratio is sufficient to obtain the line position to within $10$~MHz.  The largest contribution to the experimental uncertainty is systematic, and derives from the estimated frequency offset of $\sim150$~MHz associated with frequency chirp during the pulse amplification process in the PDA~\cite{Eikema1997}. In total, we estimate the experimental uncertainty of the transition energies to be $160$~MHz or $0.005$ \wn.

The measured transition frequencies of $Q$-, $O$- and $S$-branch lines are listed in Table~\ref{tab:transitions}. To validate the accuracy of the present measurements, we have remeasured the $Q(5)$ transition energy and obtained agreement to within $0.000\,35$ \wn\ of the previously measured value at higher accuracy ($0.000\,15$ \wn)~\cite{Salumbides2008}. This indicates that for this particular frequency range, the chirp effects in the PDA are much less than estimated. Nevertheless, we retain the conservative estimate (0.005 \wn) in view of deviations found in other wavelength ranges~\cite{Bailly2010}.

 \begin{table}
 \caption{\label{tab:transitions}
 Frequencies (in \wn) of two-photon transitions in the H$_2$ \EFX~(0,0) band.
 The uncertainty is estimated to be $0.005$ \wn.
 }
 \begin{ruledtabular}
 \begin{tabular}{rddd}
 \multicolumn{1}{c}{\textrm{$J$}}&
 \multicolumn{1}{c}{\textrm{$Q(J)$}}&
 \multicolumn{1}{c}{\textrm{$O(J)$}}&
 \multicolumn{1}{c}{\textrm{$S(J)$}}\\
 \colrule
 6		& 98\,046.299	& 		&	   \\
 7		& 97\,689.899	& 		&	   \\
 8		& 97\,291.881	& 96\,409.257	&	   \\
 9		& 96\,855.212	& 95\,875.406	&	   \\
 10		& 96\,383.125	& 95\,312.908	& 97\,610.629\\
 11		& 95\,878.434	& 94\,725.110	& 97\,175.393\\
 12		& 95\,342.943	& 94\,115.446	&	   \\
 13		& 94\,783.662	& 93\,486.710	& 96\,191.362\\
 14		&		& 92\,840.535	& 95\,656.314\\
 15		& 93\,591.330	&		&	   \\
 16		& 92\,971.330	&		&	   \\
 \end{tabular}
 \end{ruledtabular}
 \end{table}

Ground state rotational level energies $X(J)$ are obtained by subtracting the experimental $EF - X$ transition energies from the $EF,\,v=0,\,J = 2 - 12$ level energies as determined by Bailly~\etal~\cite{Bailly2010}.
For $J=13 -16$, the ground state rotational level energies are derived from the $Q$- and $O$- or $S$-branch transition frequencies. The combination differences between $Q(J)$ and $S(J-2)$ and between $Q(J)$ and $O(J+2)$ yield ground state energy splittings that validate the assignments of the transitions and the $EF(J)$ level energies of Bailly~\etal~\cite{Bailly2010}; moreover, these measurements demonstrate that $EF(13)$ was missassigned in Ref.~\cite{Bailly2010}.
The resulting values are listed in Table~\ref{tab:level_energy}, which also include level energies $X(J = 2 - 5)$ derived from Refs.~\cite{Hannemann2006, Salumbides2008} and the $X(J=1)$ level energy, signifying the ortho-para splitting quoted from Jennings~\etal~\cite{Jennings1987}, to complete the rotational sequence.

 \begin{table}
 \caption{\label{tab:level_energy}
Rotational energies $X(J)$ of H$_2$ $X,\,v=0$ levels with respect to the $X(J=0)$. The relativistic and QED corrections \XQED, with respect to \XQEDg, are obtained in combination with calculations in Ref.~\cite{Pachucki2009}\ (see text for discussion). All values are given in \wn.
 }
 \begin{ruledtabular}
 \begin{tabular}{rdl}
 \textrm{$J$}&
 \multicolumn{1}{c}{\textrm{$X(J)$}}&
 \multicolumn{1}{c}{\textrm{\XQED}}\\
 \colrule
 1		&118.48684(10) %
\footnote{From Ref.~\cite{Jennings1987}, obtained from a fit of IR quadrupole transition frequencies.}%
				&0.00174(13)\\
 2		&354.3733(2) %
\footnote{Derived from $EF-X$ transition frequencies from Refs.~ \cite{Hannemann2006,Salumbides2008,Bailly2010}.}%
				&0.0049(2)\\
 3		&705.5189(3) \footnotemark[2]		&0.0092(3)\\
 4		&1\,168.7982(2) \footnotemark[2]	&0.0157(2)\\
 5		&1\,740.1895(3) \footnotemark[2]	&0.0220(3)\\
 \colrule
 6		&2\,414.898(5)	&0.031(5)\\
 7		&3\,187.472(5)	&0.040(5)\\
 8		&4\,051.943(5)	&0.049(5)\\
 9		&5\,001.963(5)	&0.058(5)\\
 10		&6\,030.921(5)	&0.069(5)\\
 11		&7\,132.066(5)	&0.081(5)\\
 12		&8\,298.600(5)	&0.087(5)\\
 13		&9\,523.794(7)	&0.101(7)\\
 14		&10\,801.008(9)	&0.103(9)\\
 15		&12\,123.83(1)	&0.12(1)\\
 16		&13\,485.99(1)	&0.13(1)\\
 \end{tabular}
 \end{ruledtabular}
 \end{table}

\begin{figure}
\includegraphics[width=0.9\columnwidth]{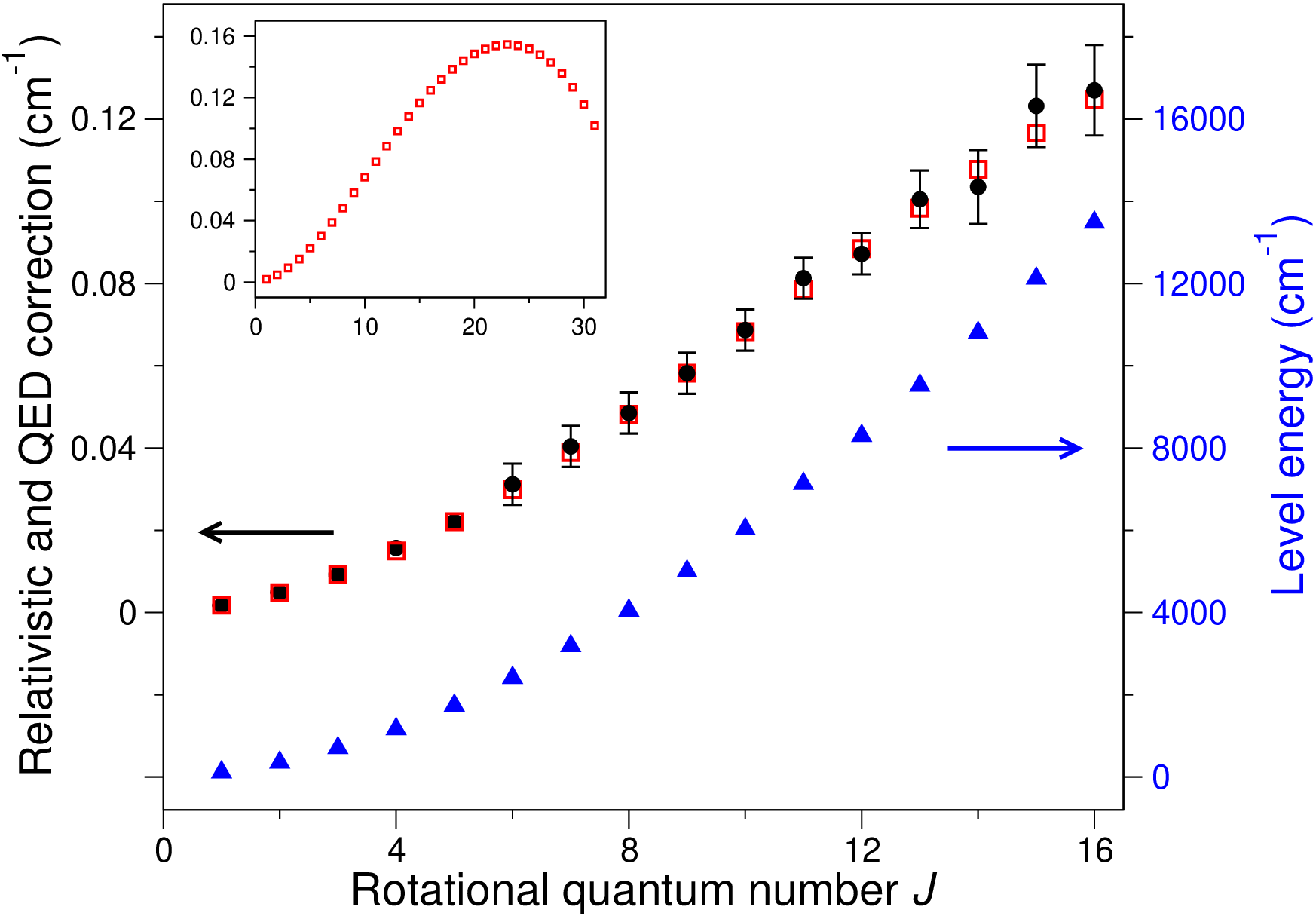}%
\caption{\label{fig:level_energy} (Color online) Rotational level energies $X(J)$ (triangles) of the \Hm\ vibronic ground state $X,\,v=0$. The relativistic and QED corrections \XQED\ (circles) are derived from the experimental $X(J)$ and the calculated values in Ref.~\cite{Pachucki2009}. The unfilled squares (also in the inset) are from the calculations of \XQED\ by Komasa~\etal~\cite{Komasa2011}.
}
\end{figure}

Relativistic and radiative corrections \XQED\ are extracted from the difference in the experimental rotational level energies $X(J)$ and those from the nonrelativistic calculations of Pachucki and Komasa \cite{Pachucki2009}. We note that the \XQED\ values, also listed in Table~\ref{tab:level_energy} are taken with respect to \XQEDg. The value \XQEDg\ for the $J=0$ quantum state is equivalent (in magnitude) to the correction for the \Hm\ dissociation energy calculated by Piszczatowski~\etal~\cite{Piszczatowski2009} to be $+0.7283(10)$ \wn. Hence, the present investigation probes \emph{differential} relativistic and radiative effects in the rotational energy sequence.
The value for \XQEDg\ from Ref.~\cite{Piszczatowski2009} comprises a full calculation of the leading-order relativistic correction and radiative corrections up to $\mathcal{O}(\alpha^3)$ orders in atomic units, and further includes estimates of radiative corrections of order $\mathcal{O}(\alpha^4)$ and recoil corrections up to $\mathcal{O}(\frac{m_\mathrm{e}}{m_\mathrm{p}}\alpha^3)$.
These result in a value of 0.5319(3) \wn\ for the leading-order relativistic correction and 0.1964(9) \wn\ for radiative corrections (including higher-order relativistic terms) for \XQEDg. In the following discussions, we refer to the combination of relativistic and radiative corrections as QED corrections.
The progression of rotational level energies $X(J)$ and the experimentally-derived QED corrections \XQED\ are plotted in Fig.~\ref{fig:level_energy}. The QED corrections are in the order of $10^{-5}$ smaller compared to the level energies, therefore, high experimental accuracy is necessary to observe the QED effects. In addition, since \XQED\ increases with increasing $J$ (for $J<22$), it is possible to observe greater (differential) relativistic and radiative corrections as higher $J$ quantum states are probed.

In Fig.~\ref{fig:level_energy} the experimentally determined values for the QED corrections are compared with the recent and yet to be published calculations of \XQED\ by Komasa~\etal~\cite{Komasa2011} for quantum states up to $J=16$. The agreement between experiment and theory for the QED corrections of $X^1\Sigma_g^+, v=0, J$ ground state rotational levels in \Hm\ is remarkably good: this is the key result of the present study. The level of agreement suggests that the systematic uncertainty contribution to the transition frequencies, due to chirp phenomena in the PDA laser system, is overestimated in the present wavelength range used. The calculated QED corrections [with respect to \XQEDg] for the full set of rotational levels (up to $J=31$) in the $X,\,v=0$ band is plotted in the inset of Fig.~\ref{fig:level_energy}.

\begin{figure}
\includegraphics[width=0.9\columnwidth]{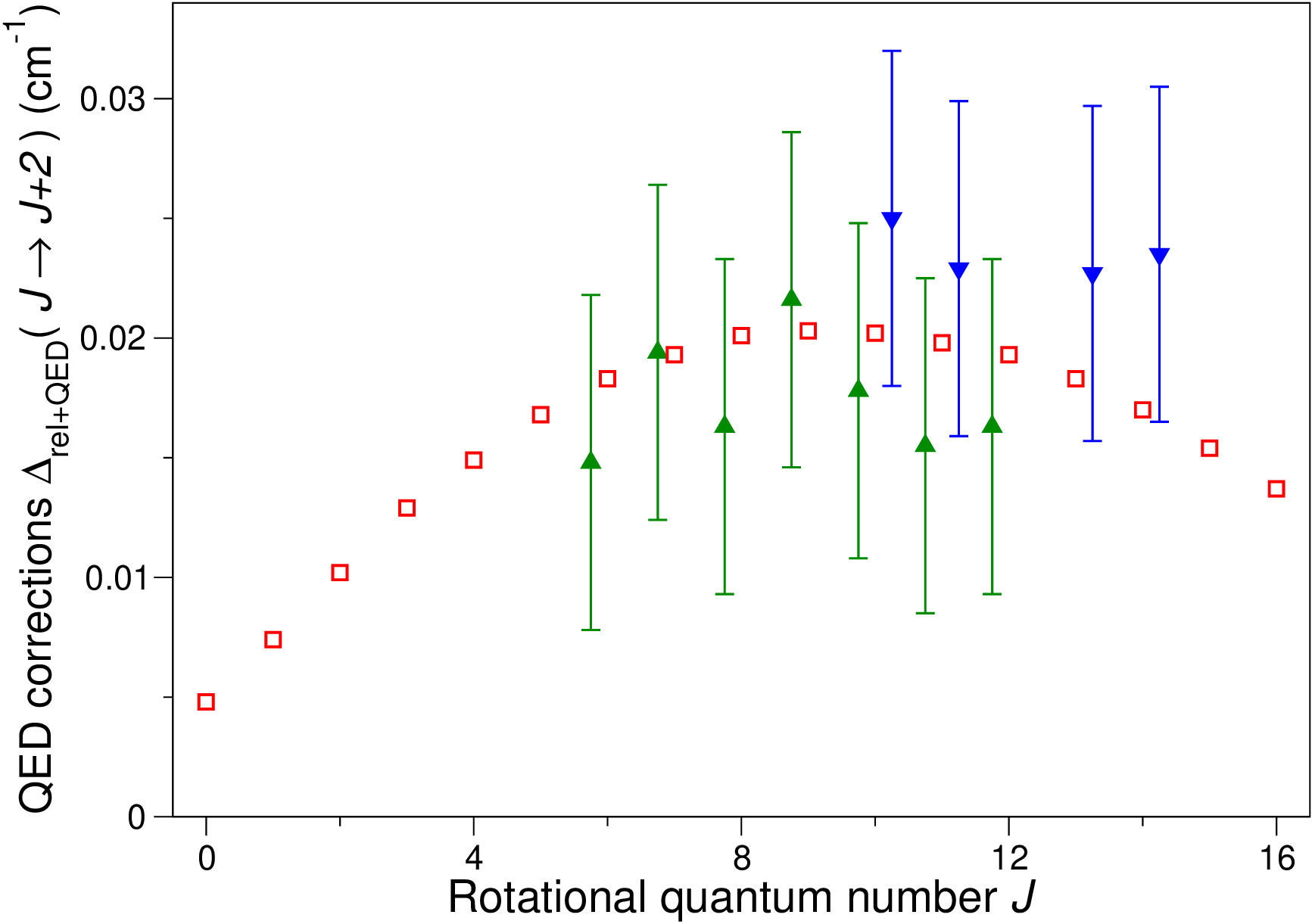}%
\caption{\label{fig:qed_level_splitting} (Color online) QED test for the combination differences in \Hm\ level energies. Upright triangles: QED contribution to combination between $O$ and $Q$ lines; inverted triangles: QED contribution to combination between $S$ and $Q$ lines. The horizontal axis of the upright and inverted triangles are shifted slightly for clarity. The unfilled squares are from the calculations of Komasa~\etal~\cite{Komasa2011}.
}
\end{figure}

The above conclusion on the QED tests in \Hm\ is in principle dependent on the correctness of the determination of $EF\,{}^1\Sigma_g^+, v=0, J$ level energies in the work by Bailly~\etal~\cite{Bailly2010}. A further test on the QED contributions relies on combination differences in the ground state, which we define as the level splittings between $J$ and $J+2$ quantum states: $\Delta(J \rightarrow J+2) = X(J+2) - X(J)$. By comparing $EF-X\,(0,0)$ transition frequencies belonging to the $O(J)$, $Q(J)$, and $S(J)$ branches, as listed in Table~\ref{tab:transitions}, values for $\Delta(J \rightarrow J+2)$ can be deduced in a straightforward manner. Again, in order to address the pure QED effects $\Delta_\mathrm{rel+QED}(J \rightarrow J+2)$ in the molecule, the nonrelativistic contributions~\cite{Pachucki2009} to the combination differences $\Delta(J \rightarrow J+2)$ are subtracted. The resulting QED contributions $\Delta_\mathrm{rel+QED}(J \rightarrow J+2)$ are finally plotted in Fig.~\ref{fig:qed_level_splitting}. The upright triangles are derived from combination differences between $O$ and $Q$ lines; the inverted triangles are derived from differences between $S$ and $Q$ lines.
These quantities do not in any way depend on actual values of $EF(J)$ excited state energies. It is noted that transitions connect only even-$J$ levels or only odd-$J$ levels in H$_2$, as is always the case in para- and orthohydrogen.
The corresponding theoretical values for the relativistic and radiative corrections $\Delta_\mathrm{rel+QED}(J \rightarrow J+2)$ for the same combination differences are also plotted in Fig.~\ref{fig:qed_level_splitting} for comparison. These purely theoretical QED contributions were derived from combining the calculations for level energies in Refs.~\cite{Pachucki2009,Komasa2011}. Figure~\ref{fig:qed_level_splitting} can be interpreted as the first derivative, with respect to $J$, of the \XQED\ sequence in Fig.~\ref{fig:level_energy}, thus explaining why the maxima in Figs.~\ref{fig:level_energy} and \ref{fig:qed_level_splitting} are located at different $J$ values.

This second test of QED, based on combination differences between $J$ and $J+2$ levels, again shows perfect agreement at the level of estimated uncertainties. While the advantage of this second QED test is that it does not rely on previously measured $EF(J)$ level energies, it has the disadvantages that significant cancellation of relativistic and radiative contributions lead to smaller energy differences, and that the uncertainties in the experimental data are slightly larger since combined errors must be taken.

With the present study, the field of molecular spectroscopy is opened up to include effects of quantum electrodynamics, i.e., the calculation of self-energy, vacuum polarization, and high-order relativistic and radiative corrections is required for an accurate representation of level energies in molecules. A rotationally-hot population distribution of \Hm\ molecules is created for a systematic and precise spectroscopic study on the sequence of rotational level energies in the \Hm\ ground electronic state. At the accuracy limit of the experiment (0.005 \wn) perfect agreement is found between experiment and theory on the QED contributions to level energies up to quantum state $J=16$. Since the accuracy level of QED calculations is claimed to be an order of magnitude more precise than the present experiments~\cite{Komasa2011}, there remains room for improvement to conduct more stringent experimental QED tests in \Hm. 

This work was financially supported by the Netherlands Foundation for Fundamental Research (FOM). We thank Dr. K. Pachucki and collaborators for making available their full set of QED calculations prior to publication.

\providecommand{\Yu}{Yu}
%

\end{document}